\documentclass[sigconf]{acmart}

\newcommand{\ie}{i.e.,\ }
\newcommand{\eg}{e.g.,\ }

\usepackage{bm}
\usepackage{amsfonts}
\usepackage{enumitem}
\AtBeginDocument{%
  }

\copyrightyear{2026}
\acmYear{2026}
\setcopyright{cc}
\setcctype{by-nc-nd}
\acmConference[WWW Companion '26]{Companion Proceedings of the ACM Web Conference 2026}{April 13--17, 2026}{Dubai, United Arab Emirates}
\acmBooktitle{Companion Proceedings of the ACM Web Conference 2026 (WWW Companion '26), April 13--17, 2026, Dubai, United Arab Emirates}
\acmPrice{}
\acmDOI{10.1145/3774905.3793106}
\acmISBN{979-8-4007-2308-7/2026/04}

\begin{document}

\title[DualMind: Agent-Based Cognitive-Affective Opinion Cascades]{\texttt{DualMind}: Towards Understanding Cognitive-Affective Cascades in Public Opinion Dissemination via Multi-Agent Simulation}

\author{Enhao Huang}
\authornote{These authors contributed equally to this research.}
\email{huangenhao@zju.edu.cn}
\orcid{0009-0008-8059-3892}
\affiliation{%
  \institution{Zhejiang University}
  \city{Hangzhou}
  \country{China}
}

\author{Tongtong Pan}
\authornotemark[1] 
\email{pantongtong@zju.edu.cn}
\orcid{0009-0008-2866-7145}
\affiliation{%
  \institution{Zhejiang University}
  \city{Hangzhou}
  \country{China}
}

\author{Shuhuai Zhang}
\authornotemark[1] 
\email{shuhuaizhang@zju.edu.cn}
\orcid{0009-0001-4737-5921}
\affiliation{%
  \institution{Zhejiang University}
  \city{Hangzhou}
  \country{China}
}

\author{Qishu Jin}
\email{brucejin@zju.edu.cn}
\orcid{0009-0008-0871-4186}
\affiliation{%
  \institution{Zhejiang University}
  \city{Hangzhou}
  \country{China}
}

\author{Liheng Zheng}
\email{lihengzheng@zju.edu.cn}
\orcid{0009-0007-1387-905X}
\affiliation{%
  \institution{Zhejiang University}
  \city{Hangzhou}
  \country{China}
}

\author{Kaichun Hu}
\email{tauhkc@zju.edu.cn}
\orcid{0009-0007-6547-6260}
\affiliation{%
  \institution{Zhejiang University}
  \city{Hangzhou}
  \country{China}
}

\author{Yiming Li}
\authornote{Corresponding author.}
\email{liyiming.tech@gmail.com}
\orcid{0000-0002-2258-265X}
\affiliation{%
  \institution{Nanyang Technological University}
  \country{Singapore}
}

\author{Zhan Qin}
\email{qinzhan@zju.edu.cn}
\orcid{0000-0001-7872-6969}
\affiliation{%
  \institution{Zhejiang University}
  \city{Hangzhou}
  \country{China}
}

\author{Kui Ren}
\email{kuiren@zju.edu.cn}
\orcid{0000-0002-1969-2591}
\affiliation{%
  \institution{Zhejiang University}
  \city{Hangzhou}
  \country{China}
}

\renewcommand{\shortauthors}{Enhao Huang et al.}

\begin{abstract}

Forecasting public opinion during PR crises is challenging, as existing frameworks often overlook the interaction between transient affective responses and persistent cognitive beliefs. To address this, we propose \texttt{DualMind}, an LLM-driven multi-agent platform designed to model this dual-component interplay. We evaluate the system on 15 real-world crises occurring post-August 2024 using social media data as ground truth. Empirical results demonstrate that \texttt{DualMind} faithfully reconstructs opinion trajectories, significantly outperforming state-of-the-art baselines. This work offers a high-fidelity tool for proactive crisis management. Code is available at \url{https://github.com/EonHao/DualMind}.
\end{abstract}

\ccsdesc[500]{Computing methodologies~Multi-agent systems}

\keywords{Public Opinion Dissemination, LLM Agent, Multi-Agent Systems, Social Simulation, Computational Social Science}


\maketitle

\section{Introduction}
\label{sec:introduction}
In the contemporary digital environment, the diffusion of public opinion across online social networks is highly influential yet difficult to anticipate. Public relations (PR) crises can escalate with exceptional speed, posing significant risks to organizational stability \cite{stabler2020corporate}. Recent high-profile cases illustrate how even single misjudgments can trigger consumer boycotts, substantial losses in market capitalization, and enduring reputational damage \cite{vosoughi2018spread}. Beyond corporate impacts, the rapid circulation of polarizing narratives can also undermine public trust and exacerbate societal tensions. Forecasting these dynamics remains a major hurdle, as they are shaped by heterogeneous user groups, intricate social structures, and platform-specific algorithmic curation. Traditional assessment methods, like surveys and focus groups, are too slow and coarse-grained for real-time discourse. Even computational and agent-based models, while faster, typically oversimplify human interaction by reducing complex, nuanced opinions to single numerical values or predefined categories \cite{Yang2024method}. This failure to model the rich semantic and emotional context of online conversations underscores an urgent need for advanced simulation tools capable of supporting proactive strategic planning and effective risk assessment \cite{aral2021hype}.

\begin{figure*}[!t]
\vspace{-1.2em}
\centering
\includegraphics[width=\linewidth]{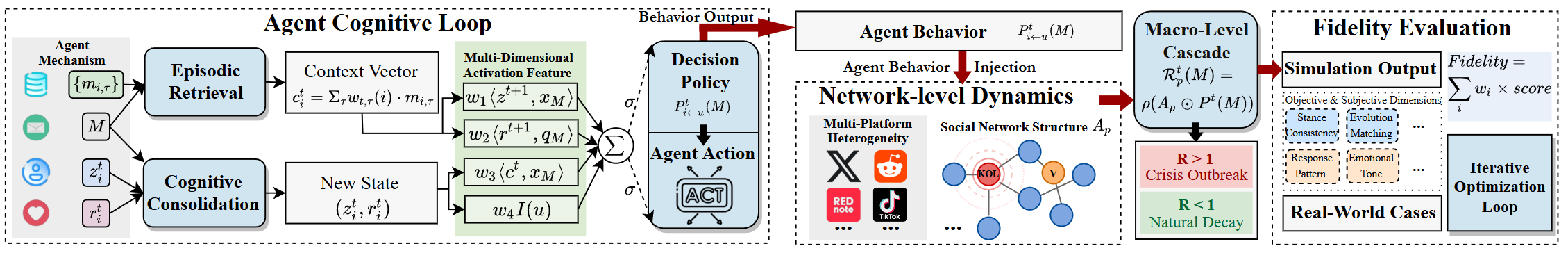}
\vspace{-2.5em}
\caption{Overview of the \texttt{DualMind} pipeline.}
\label{fig:framework}
\vspace{-1em}
\end{figure*}

The advent of Large Language Models (LLMs) has accelerated progress in computational social science, leading to the widespread application of LLM-based autonomous agents in research \cite{NEURIPS2024_1cb57fcf}. In particular, a growing body of work demonstrates that these agents can effectively simulate human social behaviors \cite{cau2025language}. When instantiated within simulated environments, LLM-based agents serve as viable \emph{computational subjects}, capable of replacing or augmenting human participants in scientific studies and social simulations \cite{sun2025llm}. This agent-based simulation approach offers a novel and scalable avenue for exploring complex societal dynamics, such as opinion formation and information diffusion, in a controlled yet realistic manner.

Despite this potential, existing simulations often lack the specific architecture for high-stakes PR crises. Current LLM-based agent studies typically focus on general belief evolution through prompt-engineered biases \cite{chuang-etal-2024-simulating} or use hybrid models where agents only handle specific tasks like information alteration \cite{hu2024llm}. While these advance beyond traditional numerical models that reduce opinions to single scores \cite{Yang2024method}, a critical gap remains. Specifically, existing platforms fail to model the interplay between rapid, fluctuating affective states and the slower, more stable evolution of cognitive beliefs. To bridge this gap, we propose \texttt{DualMind}, a novel multi-agent simulation platform centered on this dual-component architecture. Our system constructs a virtual social network populated by a large cohort of LLM-based agents, each endowed with a distinct persona, a memory module for cognitive continuity, and platform-specific behavioral protocols. These agents perceive information, interact with one another, and express opinions according to their unique characteristics and simulated platform constraints (\eg Twitter, Weibo, Reddit), thereby creating a high-fidelity digital microcosm for studying crisis dynamics.

To demonstrate the validity and utility of our platform, we conduct a rigorous evaluation grounded in real-world events. We employ a set of foundational LLMs whose knowledge bases are verifiably truncated before August, 2024 (including both open-source and proprietary models) to prevent data contamination. We then simulate the public response to 15 high-profile PR crises that occurred after August, 2024 across diverse geopolitical regions, including China, the United States, and Europe. Our experimental results reveal a strong correspondence between the simulated opinion trajectories generated by \texttt{DualMind} and the actual, documented evolution of public sentiment in these historical cases. This validation underscores the platform's potential as a reliable tool for both academic research and practical PR strategy testing.

\section{\texttt{DualMind}: A Multi-Agent Simulation System}
\label{sec:framework}

\subsection{Methodology}

In general, \texttt{DualMind} formalizes public-opinion dynamics as the coupled evolution of cognitive states inside LLM-driven agents and content-sensitive, emotion-aware diffusion over platform-specific networks (see Fig.~\ref{fig:framework}). Each agent $i$ maintains a dual latent state composed of a slowly evolving \emph{semantic persona} $\bm{z}_i^t\in\mathbb{R}^d$ and a rapidly fluctuating \emph{affective state} $\bm{r}_i^t\in\mathbb{R}^K$, complemented by an episodic store of past interactions $\{\bm{m}_{i,\tau}\}$ annotated with content embeddings $\bm{x}_{i,\tau}\in\mathbb{R}^d$ and emotion vectors. When a candidate message $M$ with embedding $\bm{x}_M$ and emotion $\bm{q}_M$ arrives at time $t$, the agent retrieves a context vector by attention-weighted, recency-decayed aggregation over its episodic store:

\vspace{-1.25em}
\begin{equation}
\label{eq:episodic}
\bm{c}_i^t \;=\; \sum_{\tau < t} \underbrace{\frac{\exp\!\big\{\beta\,\langle \bm{x}_{i,\tau}, \bm{x}_M\rangle\big\}\,\delta^{\,t-\tau}}{\sum_{\ell < t}\exp\!\big\{\beta\,\langle \bm{x}_{i,\ell}, \bm{x}_M\rangle\big\}\,\delta^{\,t-\ell}}}_{w_{t,\tau}^{(i)}}
\,\bm{m}_{i,\tau},
\end{equation}

where $\beta>0$ controls semantic selectivity, $\delta\in(0,1)$ implements exponential recency, and $w_{t,\tau}^{(i)}$ are normalized retrieval weights. This retrieval realizes RAG within the agent by ensuring that decision-making is conditioned on semantically relevant and temporally proximate experiences rather than on a fixed context window.

The dual-stream cognitive state updates in response to $M$ through a gated, coupled rule that preserves slow–fast separation between long-term beliefs and transient emotions. Let $\sigma(x)=(1+e^{-x})^{-1}$ be the nonlinear logistic mapping, $\Pi(\bm{v})=\bm{v}-\langle \bm{v},\bm{z}_i^t\rangle \bm{z}_i^t$ is the projection onto the tangent space at $\bm{z}_i^t$ (so that beliefs drift along the unit sphere); with agent-specific emotional persistence $\eta\in(0,1)$, adaptation rate $\gamma>0$, and affective gate gain $\alpha>0$, we write
\begin{equation}
\label{eq:dualupdate}
\begin{bmatrix}
\bm{z}_i^{t+1}\\[2pt]
\bm{r}_i^{t+1}
\end{bmatrix}
=
\begin{bmatrix}
\bm{z}_i^{t}\\[2pt]
\eta\,\bm{r}_i^{t}
\end{bmatrix}
+\gamma\, \sigma\!\big(\alpha\, \langle \bm{r}_i^{t}, \bm{q}_M\rangle\big)\!
\begin{bmatrix}
\Pi\!\big(\bm{x}_M - \langle \bm{x}_M,\bm{z}_i^{t}\rangle \bm{z}_i^{t}\big)\\[2pt]
\bm{q}_M - \bm{r}_i^{t}
\end{bmatrix},
\end{equation}
followed by re-normalizing $\bm{z}_i^{t+1} \leftarrow \bm{z}_i^{t+1} / \|\bm{z}_i^{t+1}\|_2$. In general, Eq.~\eqref{eq:dualupdate} formalizes a psychologically grounded consolidation mechanism: only when the incoming emotion resonates with the current affect does the long-term persona take a small geodesic step toward the message content; otherwise, the update remains largely confined to the affective channel.

The decision-making of each agent is further modeled as a probabilistic discrete-choice policy, dubbed \emph{Polarized Affective Cascade Model} (PAACM), that integrates long-term alignment, affective resonance, and episodic coherence, together with social authority and platform norms. Let $I(u)$ denote the calibrated source influence of the sender $u$ (\eg follower-weighted reputation), $\theta_i$ the intrinsic activation threshold of agent $i$, and $b_{p(i)}$ a platform bias term determined by the platform hosting $i$. The activation probability that $i$ engages with $M$ received from $u$ at time $t$ under the PAACM is
\begin{equation}
\label{eq:activation}
P_{i\leftarrow u}^{t}(M)
=\sigma\!\Big(
w_1\,\langle \bm{z}_i^{t}, \bm{x}_M\rangle
+w_2\,\langle \bm{r}_i^{t}, \bm{q}_M\rangle
+w_3\,\langle \bm{c}_i^{t}, \bm{x}_M\rangle
+w_4\,I(u)+ b_{p(i)} - \theta_i
\Big),
\end{equation}
with nonnegative weights $\{w_i\}_{i=1}^4$ learned or tuned per-platform. 

The formulation in Eq.~\eqref{eq:activation} subsumes the classic Independent Cascade model \cite{kempe2005influential} as a special case, which is recovered by setting $w_1=w_2=w_3=0$ and $w_4$ to a constant. Our generalized model extends this framework by extending the propagation process to incorporate content semantics, affect, and memory.

The foregoing micro-level agent policy will also induce macro-level opinion diffusion whose explosiveness can be characterized spectrally \cite{wang2003epidemic}. Specifically, writing $\bm{A}_{p}\in\{0,1\}^{n\times n}$ for the directed adjacency on platform $p$ and $\bm{P}^{t}(M)\in[0,1]^{n\times n}$ for the matrix with entries $[\bm{P}^{t}(M)]_{i,u}=P_{i\leftarrow u}^{t}(M)$, the expected reproduction coefficient of message $M$ at time $t$ can be formulated as
\begin{equation}
\label{eq:repro}
\mathcal{R}_{p}^{t}(M)\;=\;\varrho\!\left(\bm{A}_{p}\,\odot\,\bm{P}^{t}(M)\right),
\end{equation}
where $\odot$ denotes the Hadamard product and $\varrho(\cdot)$ the spectral radius. Supercritical cascades satisfy $\mathcal{R}_{p}^{t}(M)>1$, leading to self-sustaining opinion waves; subcritical regimes attenuate. This spectral control parameter provides a principled, platform-aware knob for intervention design: PR strategies generated by the LLM planner are accepted when they reduce $\mathcal{R}_{p}^{t}(M)$ below unity without degrading alignment with desired narrative constraints.

\subsection{System Architecture}

The system employs a decoupled front-end/back-end architecture, comprising a React front-end, a FastAPI back-end, and an agent layer based on LangChain and LangGraph.

The \textbf{front-end} utilizes a React (v19.1.1) and TypeScript stack with Ant Design.Social network visualization leverages the native Canvas API, chosen over libraries like D3.js for high-performance dynamic propagation effects.

The \textbf{back-end} simulator runs in a Python environment, employing the high-performance asynchronous framework FastAPI (over traditional Flask or Django) to serve RESTful APIs via Uvicorn.

The core \textbf{agent layer} utilizes LangChain and LangGraph to establish a structured cognitive workflow. Agent execution follows a controlled concurrency model: active agents execute sequentially in random order per round to avoid API concurrency limits. Besides, we use a multi-model LLM strategy: the lightweight \texttt{gpt-4o-mini} handles high-frequency agent tasks (\eg generating posts, evaluating stances), while the more powerful \texttt{gemini-1.5-pro} is used for post-hoc analysis, generating the final fidelity reports.

\section{Demonstration and Case Studies}

To validate the effectiveness of \texttt{DualMind} in simulating complex public relations (PR) crisis dynamics and to demonstrate its core functionalities, we designed and implemented an interactive demonstration system. The core of this system is its \textbf{high-fidelity public opinion simulation capability}, which is built upon a validated cognitive-affective agent architecture. This system supports two core application scenarios, focusing respectively on analyzing past events and simulating future strategies.

\subsection{Demonstration}
\subsubsection{Scenario 1: Retrospective Analysis of Real-World Cases}
\begin{figure}[!t]
  \centering
  \vspace{-1em}
  \includegraphics[width=\linewidth]{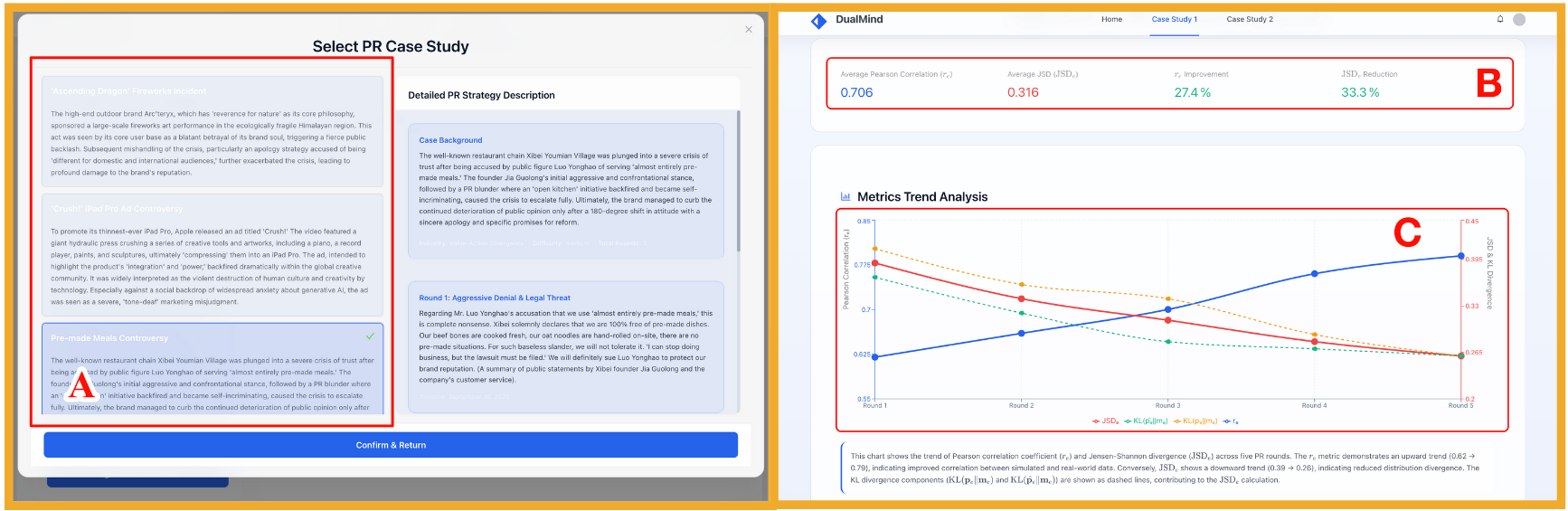}
  \vspace{-2em}
  \caption{User interface of Scenario 1 (Real-world Cases). (A) The user selects a real-world event from the "Select PR Case Study" window to set the ground truth; (B) The "Comparative Analysis Report" provides immediate, top-level fidelity metrics (\eg overall Pearson's $r$ and JSD); (C) The "Metrics Trend Analysis" graph allows the user to assess the fidelity of the opinion trajectory as it evolves over simulation rounds.}
  \label{fig:scenario1_demo}
\vspace{-6pt}
\end{figure}

This scenario focuses on \textbf{analyzing and reproducing past, real-world events} to help users understand how public opinion evolved and how interventions shaped the final outcome (see Fig.~\ref{fig:scenario1_demo}).

The system first presents a \textbf{Select PR Case Study} window. The user selects from 15 representative corporate PR crises curated from public data streams across major social media platforms in the U.S., China, and Europe. All cases occurred after August 2024 and were intentionally chosen beyond mainstream LLM knowledge cutoffs to avoid prior exposure to outcomes.

Upon selection, the user sees the detailed background and real strategy nodes. After confirmation, the simulation starts automatically and follows an authentic day-level timeline. The system ingests each critical event (e.g., the Day-1 official statement) to drive the agent cohort’s simulation. After each node, the user can proceed to the next event or click \textbf{Generate Report}.

The system then presents a \textbf{Comparative Analysis Report}. Similarity scores, Pearson’s correlation coefficient ($r$) and Jensen--Shannon Divergence (JSD), summarize proximity to the Ground Truth. The user can further browse the \textbf{Metrics Trend Analysis} graph to assess fidelity at the trajectory level.

\subsubsection{Scenario 2: User-specified Simulation}
\begin{figure}[t]
  \centering
  \vspace{-1em}
  \includegraphics[width=\linewidth]{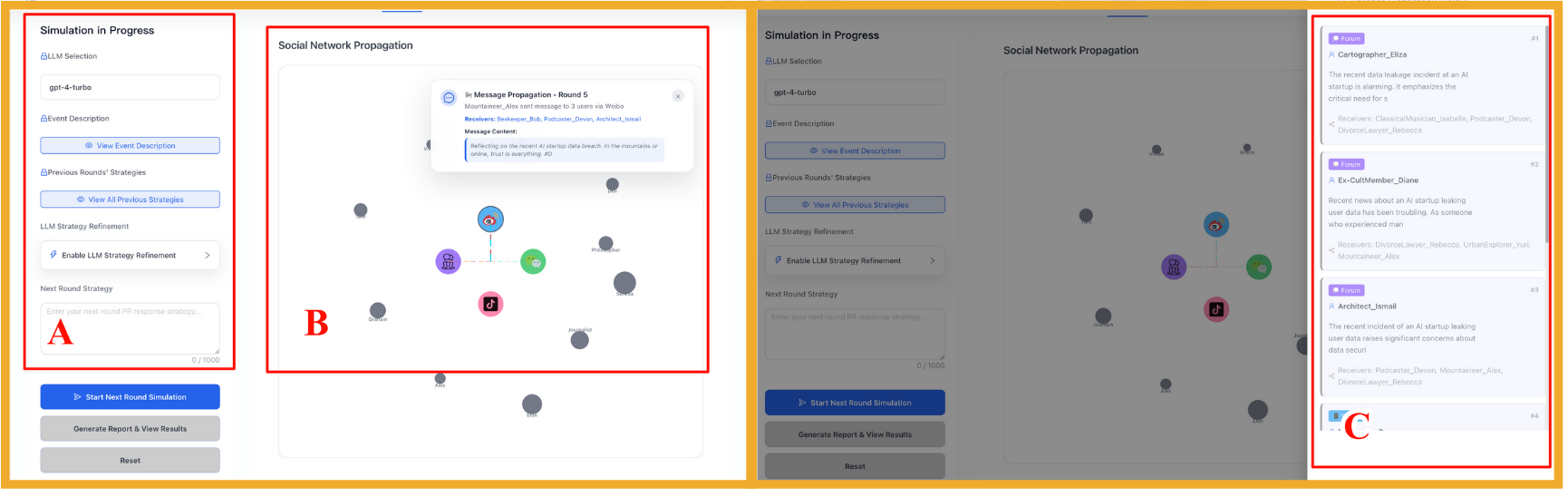}
    \vspace{-2em}
  \caption{User interface for Scenario 2. (A) The user iteratively inputs natural language PR strategies in the left panel; (B) The center panel provides a real-time visualization of the opinion propagation network; (C) The right ``Message Propagation History" panel tracks agent posts. Users analyze feedback from (C) to refine strategies in (A) and observe the new strategy's diffusion in (B).}
  \Description{Screenshots of the Case Study 2 interface, including the network graph for propagation and the side panel for strategy input.}
  \label{fig:scenario2_demo}
\vspace{-12pt}
\end{figure}

This scenario targets user-specified PR events and functions as an interactive \textbf{strategy rehearsal sandbox} that generates actionable recommendations to support decision-making (see Fig.~\ref{fig:scenario2_demo}).

The user can conduct a custom simulation. The user defines an event in the \textbf{Event Description} text box and inputs the first intervention strategy in the \textbf{Next Round Strategy} field. After clicking \textbf{Start Next Round Simulation}, the system samples agents from an integrated persona library aligned with real-world demographic statistics and instantiates a dynamic social network of heterogeneous actors, enabling real-time visualization of opinion diffusion.

The \textbf{Message Propagation History} sidebar provides a step-by-step trace of cross-platform message spread through actors, including amplification by Key Opinion Leaders. After each round, the user reviews opinion dynamics and revises \textbf{Next Round Strategy}.

At any time, clicking \textbf{Generate Report \& View Results} generates an analytical report—including stance trajectories and topic trends—allowing the user to evaluate cumulative impacts and identify strategies most effective for the desired PR objectives.

\subsection{Case Study and Evaluation}
\label{sec:experiments}

We hereby perform quantitative evaluations to assess \texttt{DualMind}'s ability to faithfully replicate both the temporal evolution of public opinion (\ie opinion trajectories) and the eventual outcomes (\ie aggregate stance distributions) of real-world PR crises, using real-world data as ground truth.

\begin{figure}[!t]
\vspace{-1.2em}
\centering

\includegraphics[width=0.98\columnwidth,trim=0 6mm 0 6mm, clip]{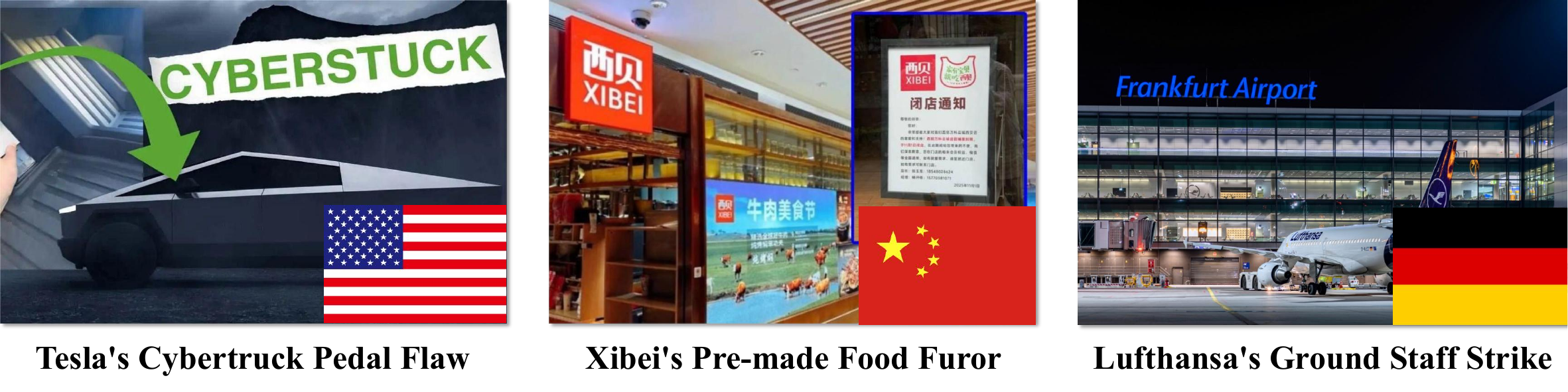}
\vspace{-1.2em}
\caption{Some major public relations event cases used in assessments from around the world.}
\vspace{-2em}
\label{fig:cases}
\end{figure}

\subsubsection{Baselines and Settings}
\label{subsec:setup}

We evaluate our system on a curated dataset of 15 real-world PR crises (late 2024–2025), evenly distributed across the United States (5), China (5), and Europe (5) (see Fig.~\ref{fig:cases}). All models are instantiated using LLMs with knowledge cutoffs prior to August 2024 to avoid data contamination. For each crisis, we simulate 100 agents that represent heterogeneous social media users interacting over a crisis-specific network, and we apply the same agent population size and interaction protocol to all baselines for a fair comparison. We benchmark \texttt{DualMind} against three state-of-the-art (SOTA) baselines, including:
\begin{itemize}[leftmargin=1.2em,itemsep=2pt,topsep=2pt]
    \item \textbf{LAID} \cite{hu2024llm}: An LLM-enhanced agent model for information propagation, evaluated on four hypothetical scenarios (\eg viral marketing) rather than real-world crises.
    \item \textbf{LPOD} \cite{Yang2024method}:
    A non-LLM agent-based model that treats agents as social media users and updates their opinions and network links through equation-driven rules. It incorporates event spreading, link prediction, and trust estimation, and demonstrates high predictive accuracy on both synthetic and real Weibo datasets.
    \item \textbf{LLM-GA} \cite{chuang-etal-2024-simulating}:
    A framework that simulates opinion dynamics using networks of LLM-driven agents engaging in natural-language interactions. It evaluates 15 topics with known ground truth and identifies a systematic truth-convergence tendency in LLM agents during belief updates.
\end{itemize}
All simulated outputs, $\hat{\bm{y}}_e$ and $\hat{\bm{p}}_e$, are generated using the same protocol across models and averaged over 5 stochastic seeds.

\subsubsection{Metrics}
\label{subsec:metrics}

We measure \textbf{process fidelity} using Pearson's correlation coefficient $r_e$ between the empirical trajectory $\bm{y}_e$ and the simulated trajectory $\hat{\bm{y}}_e$; higher $r_e$ indicates better temporal alignment. We measure \textbf{outcome fidelity} using the Jensen--Shannon Divergence $\mathrm{JSD}_e(\bm{p}_e \Vert \hat{\bm{p}}_e)$ between the final stance distributions, where lower $\mathrm{JSD}_e$ indicates a closer match.

\subsubsection{Results}
\label{subsec:results}

\begin{figure}[!t]
\centering
\vspace{-1.4em}
\includegraphics[width=0.9\columnwidth,trim=0 5mm 0 5mm, clip]{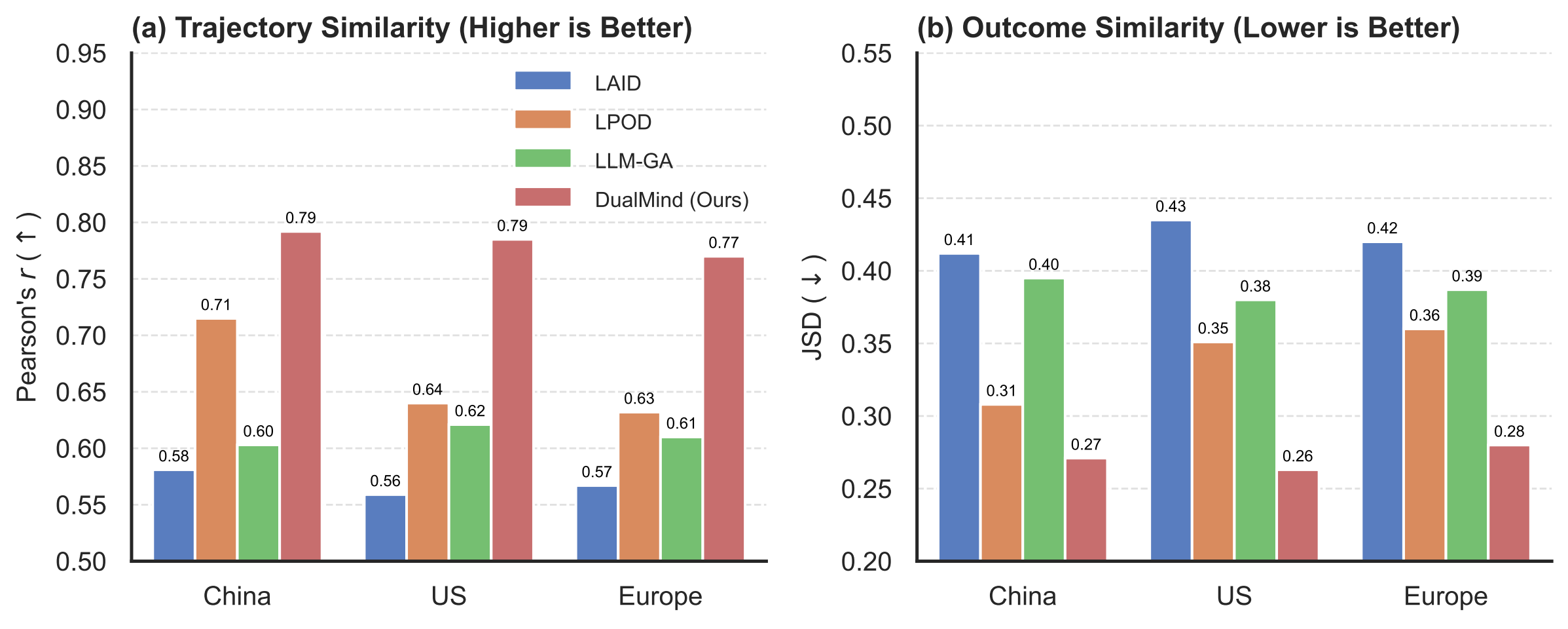}
\vspace{-1.4em}
\caption{Fidelity comparison. \texttt{DualMind} achieves the highest trajectory similarity and the lowest outcome divergence.}
\label{fig:main_results}
\vspace{-1.2em}
\end{figure}

As visualized in Figure~\ref{fig:main_results}, \texttt{DualMind} consistently and significantly outperforms all baselines across all 15 cases. Our model achieves the highest average trajectory similarity (overall $\overline{r}{\approx}0.78$) and the lowest average outcome divergence (overall $\overline{\mathrm{JSD}}{\approx}0.27$). 

This result highlights two key findings. Firstly, the model demonstrates strong cross-cultural robustness, showing high-fidelity performance across the distinct media environments of US, China, and Europe. Secondly, \texttt{DualMind} shows clear superiority over SOTA models, significantly outperforming even the strong LPOD baseline \cite{Yang2024method} in its own validated domain.

In particular, whereas the LLM-GA framework \cite{chuang-etal-2024-simulating} reported a persistent `truth-bias' that drives agents toward factual consensus, our real-world PR case studies demonstrate that \texttt{DualMind} overcomes this structural limitation. By integrating cognitive–affective cascades with realistic multi-agent social simulations, it faithfully captures the complex and fact-resistant opinion dynamics observed in human social networks.

\section{Conclusion and Future Works}

In this demo, We presented \texttt{DualMind}, an LLM-driven multi-agent simulator for cognitive-affective opinion cascades in PR crises. On real-world crises after August 2024, \texttt{DualMind} demonstrates high fidelity in reproducing both opinion trajectories and stance distributions, outperforming SOTA baselines. Limitations arise from simplified cognition, abstracted platform algorithms and network scales. Next, we will enrich agent reasoning, incorporate external information streams, and use the platform to automatically explore and evaluate PR intervention strategies.

\section*{Ethical Use of Data and Informed Consent}

This study relies solely on publicly available data for analysis. No human subjects were involved, and informed consent was not required. The work complies with all applicable ACM policies.

\begin{acks}
This research is supported in part by the ``Pioneer'' and ``Leading Goose'' R\&D Program of Zhejiang (2024C01169), the Kunpeng-Ascend Science and Education Innovation Excellence/Incubation Center, and the National Natural Science Foundation of China under Grants (625B1032, 62441238, U2441240).
\end{acks}

\bibliographystyle{ACM-Reference-Format}
\bibliography{sample-base}


\end{document}